\documentclass[12pt]{article}

\pdfoutput = 1
\usepackage{graphicx}
\graphicspath{{Graphs/}}
\usepackage{titlesec}
\titlelabel{\thetitle.\quad}
\begin{document}
\date{Today}
\title{{\bf{\Large Scalar-metric quantum cosmology with Chaplygin gas and perfect fluid  }}}

\author{
{\bf {\normalsize Saumya Ghosh}$^{a}
$\thanks{sgsgsaumya@gmail.com, sg14ip041@iiserkol.ac.in}},
{\bf {\normalsize Sunandan Gangopadhyay}
$^{a}$\thanks{sunandan.gangopadhyay@gmail.com, sunandan@iiserkol.ac.in, sunandan@associates.iucaa.in}},
{\bf {\normalsize Prasanta K. Panigrahi}$^{a}
$\thanks{pprasanta@iiserkol.ac.in}}\\
$^{a}$ {\normalsize Indian Institute of Science Education and Research Kolkata}\\
{\normalsize Mohanpur 741246, Nadia, West Bengal, India }\\
}
\date{}

\maketitle

\begin{abstract}
{\noindent In this paper we consider the flat FRW cosmology with a scalar field coupled with the metric along with generalized Chaplygin gas and perfect fluid comprising the matter sector. We use the Schutz's formalism to deal with the generalized Chaplygin gas sector. The full theory is then quantized canonically using the Wheeler-DeWitt Hamiltonian formalism. We then solve the WD equation with appropriate boundary conditions. Then by defining a proper completeness relation for the self-adjointness of the WD equation we arrive at the wave packet for the universe. It is observed that the peak in the probability density gets affected due to both fluids in the matter sector, namely, the Chaplygin gas and perfect fluid.
  }
\end{abstract}
\vskip 1cm



\section{Introduction}
The Chaplygin gas model has proved to be an interesting model due to its ability to describe the accelerated expansion of our universe \cite{Pedram - 9}. The generalized version of this model \cite{Pedram - 9, Pedram - 10} has also been studied extensively \cite{Pedram - 23, Pedram - 24}. Interestingly it was found that this model emerges from the Nambu-Goto action of relativistic strings in the light cone coordinates \cite{Pedram - 26}.
Quantum mechanical analysis of an FRW cosmolgical model with generalized Chaplygin gas was discussed in \cite{Pedram - 30, Pedram - 31}. Such an analysis is important in its own right because the universe must surely be governed quantum mechanically when the linear size of the universe was very close to the Plank scale($10^{-33}$cm). Further one can make a positive expectation that a quantum mechanical description of our universe may result in a singularity free birth of the universe.
In \cite{Pedram et. al}, apart from the Chaplygin gas coupled to gravity, the perfect fluid was also included. The early and late time behavior of the expectation of the scale factor was then obtained analytically using the Schutz formalism \cite{B. Schutz 01, B. Schutz 02}. This formalism has also been used extensively in \cite{Pedram et. al, Barun - 21, Barun - 22, B. Majumder}. In \cite{B. Majumder} the quantum dynamics of the spatially flat FRW model with Chaplygin gas and a scalar field coupled to the metric has been investigated. The wave-packet was found from the linear superposition of the wave-functions of the Schrodinger-Wheeler-DeWitt(SWDW) equation. It was found to show two distinct peaks. Similar studies has also been carried out in various other cosmological models \cite{fuchka}-\cite{Pal2}. 

In this paper, motivated by the works discussed above \cite{Pedram et. al, B. Majumder}, we study the quantum dynamics of the FRW model with Chaplygin gas, scalar field coupled to the metric together with the inclusion of the perfect fluid. We apply the Schutz formalism together with the canonical approach \cite{Lapchinskii et. al} to obtain the super Hamiltonian. This is then quantized canonically to get the SWDW equation. This is then solved to obtain the quantum cosmological wave functions of the universe which in turn is used to construct the wave packet. 

This paper is organized as follows. In the section 2 we have discussed  the basic set up for the canonical quantization of a gravity model and also the Schutz's formalism for dealing with the matter sector.
In section 3 we have carried out the quantization of the scalar metric theory of gravity in the presence of a Chaplygin gas and perfect fluid. 
We conclude  in section 4.

\section{Basic set up for quantization}
We start by writing down the action which includes gravity with a scalar field coupled to it together with the perfect fluid and the generalized Chaplygin gas representing the matter sector 
\begin{eqnarray}
\label{2}
S&=&\int_{M}d^4x\sqrt{-g}\left[ R-F(\phi)g^{\mu\nu}\phi_{,\mu}\phi_{,\nu}\right]+2\int_{\partial M}d^3x\sqrt{h}h_{ij}K^{ij}+\int_{M}d^4x\sqrt{-g}(p_f+p_c)\\ \nonumber
&\equiv& S_g+S_m
\end{eqnarray}
where $K^{ij}$ is the extrinsic curvature tensor, $h_{ij}$ is the induced metric on the time-like hypersurface and $F(\phi)$ is an arbitrary function of the scalar field. The matter sector consists of two fluids, namely, the perfect fluid with pressure $p_f=\omega\rho_f$ and the Chaplygin gas with pressure
\begin{equation}
p_c=-\frac{A}{\rho^\alpha_c}~.
\end{equation} 
In the subsequent discussion we shall follow the Schutz formalism \cite{B. Schutz 01} to deal with only the Chaplygin gas of the matter sector. In this formalism, the four velocity of the fluid can be written down in terms of four potentials $h, \epsilon, \theta$ and $S$ as 
\begin{eqnarray}
\label{23}
u_\nu=\frac{1}{h}(\epsilon_{,\nu}+\theta S_{,\nu})
\end{eqnarray}
where $h$ and $S$ are the specific enthalpy and specific entropy respectively. The other two ($\theta$ and $\epsilon$) do not have any physical significance. 
The normalization condition of the four velocity reads
\begin{eqnarray}
\label{24}
u_\nu u^\nu=-1 ~.
\end{eqnarray}
In this paper, we shall work with the flat FRW $(k=0)$ metric given by 
\begin{eqnarray}
\label{25}
ds^2=-N^2(t)dt^2+a^2(t)[dr^2+r^2(d\vartheta^2+Sin^2\vartheta d\varphi^2)]
\end{eqnarray}
where $N(t)$ is the lapse function and $a(t)$ is the scale factor.
The Ricci scalar $R$ for this metric is given by 
\begin{eqnarray}
\label{26}
R=\frac{1}{N^3a^2}[-6a\dot{a}\dot{N}+6N{\dot{a}}^2+6Na\ddot{a}]~.
\end{eqnarray}
The gravity part of the action can now be evaluated and after dropping the surface terms reads
\begin{eqnarray}
S_g=\int dt\left[-6\frac{\dot a^2 a}{N}+\frac{1}{N}F(\phi)a^3\dot\phi^2\right]~.
\end{eqnarray}
The Hamiltonian for this action therefore reads (upto a factor of $N$)
\begin{eqnarray}
H_g=-\frac{p^2_a}{24a}+\frac{1}{4F(\phi)}\frac{p^2_\phi}{a^3}
\end{eqnarray}
where $p_a = -\frac{\partial L}{\partial \dot{a}}$ and $p_\phi = -\frac{\partial L}{\partial \dot{\phi}}$ are the canonically conjugate momenta corresponding to $a$ and $\phi$.
To proceed further, we shall now concentrate on the Chaplygin gas part of the matter sector. We shall first write down the basic thermodynamic relations following from the thermodynamic description in \cite{Lapchinskii et. al}. These read
\begin{eqnarray}
\label{27}
\rho_c=\rho_0(1+\Pi),~~~ h=1+\Pi+p_c/\rho_0,~~~ \tau dS=d\Pi+p_cd(1/\rho_0)
\end{eqnarray}
where $\rho_c$ is the total mass energy density, $\tau$ is the temperature, $\rho_0$ is the rest mass density and $\Pi$ is the specific internal energy. 
The next step is to obtain an expression for $p_c$ in terms of $h$ and $S$. This can be done with the help of the thermodynamic relations  above and reads \cite{Pedram et. al}
\begin{eqnarray}
\label{28}
p_c=-A\left[\frac{1}{A}\left(1-\frac{h^\frac{1+\alpha}{\alpha}}{S^\frac{1}{\alpha}}\right)\right]^\frac{1+\alpha}{\alpha}~.
\end{eqnarray} 
Now using eq.(s) (\ref{23}) and (\ref{24}), we get 
\begin{eqnarray}
\label{29}
p_c=-A\left[\frac{1}{A}\left(1-\frac{(\dot\epsilon+\theta\dot S)^\frac{1+\alpha}{\alpha}}{N^{\frac{1+\alpha}{\alpha}}S^\frac{1}{\alpha}}\right)\right]^\frac{1+\alpha}{\alpha}~.
\end{eqnarray}
Hence the Chaplygin part of the action reads
\begin{eqnarray}
S_c=-\int dt N a^3 A\left[\frac{1}{A}\left(1-\frac{(\dot\epsilon+\theta\dot S)^\frac{1+\alpha}{\alpha}}{N^{\frac{1+\alpha}{\alpha}}S^\frac{1}{\alpha}}\right)\right]^\frac{1+\alpha}{\alpha}~.
\end{eqnarray}

\noindent Combining the gravity sector of the action with the matter sector of the action, we now have
\begin{eqnarray}
\label{2b}
S=\int dt\left[-6\frac{\dot a^2 a}{N}+\frac{1}{N}F(\phi)a^3\dot\phi^2+Na^3\omega\rho_f-Na^3A\left[\frac{1}{A}\left(1-\frac{(\dot\epsilon+\theta\dot S)^\frac{1+\alpha}{\alpha}}{N^{\frac{1+\alpha}{\alpha}}S^\frac{1}{\alpha}}\right)\right]^\frac{1+\alpha}{\alpha}\right]~.
\end{eqnarray}
The Hamiltonian for the matter sector of the action reads
\begin{eqnarray}
H_m=-a^3{\omega}\rho_f+(Sp^{1+\alpha}_\epsilon+Aa^{3(1+\alpha)})^{\frac{1}{1+\alpha}}~.
\end{eqnarray}
We now make an assumption that we shall be considering the early universe case. Hence we assume
\begin{eqnarray}
\label{2d}
Sp^{1+\alpha}_\epsilon>>Aa^{3(1+\alpha)}~.
\end{eqnarray}
Therefore Hamiltonian for the matter sector takes the form (upto a factor of $N$)
\begin{eqnarray}
H_m=-a^3\omega\rho_f+S^{\frac{1}{1+\alpha}}p_\epsilon~.
\end{eqnarray} 
The super-Hamiltonian for the full theory now reads
\begin{eqnarray}
\label{2e}
H &=& H_g +H_m \\
&=& -\frac{p^2_a}{24a}+\frac{1}{4F(\phi)}\frac{p^2_\phi}{a^3}-a^3\omega\rho_f+S^{\frac{1}{1+\alpha}}p_\epsilon~.
\end{eqnarray}
To simplify this Hamiltonian further one uses the canonical transformations 
\begin{eqnarray}
T&=&-(1+\alpha)p_\epsilon^{-1}S^{\frac{\alpha}{1+\alpha}} p_S          \\
p_T&=&S^{\frac{\alpha}{1+\alpha}} p_\epsilon
\end{eqnarray}
along with the explicit form of the perfect fluid energy density $\rho_f =\frac{B}{a^{3(1+\omega)}}$. This leads to
\begin{eqnarray}
H=-\frac{p^2_a}{24a}+\frac{1}{4F(\phi)}\frac{p^2_\phi}{a^3}-\omega B a^{-3\omega}+p_T
\end{eqnarray}
where $p_T$ is the canonical variable corresponding to the matter sector  along the direction of the cosmic time.


\section{Quantization of the scalar-metric cosmology}
We now proceed to write down the Wheeler-De Witt (WD) equation for the Hamiltonian written down in the earlier section. For this we replace $p_a=-i\frac{\partial}{\partial a}, p_T = -i\frac{\partial}{\partial T}$ and $p_\phi=-i\frac{\partial}{\partial \phi}$ in the Hamiltonian and assume that the Hamiltonian $\hat{H}$ annihilates the wave function. This leads to the WD equation which reads (setting $\hbar=1$)  
\begin{eqnarray}
\label{4a}
\left[\frac{\partial^2}{\partial a^2}-\frac{6}{F(\phi)a^2}\frac{\partial^2}{\partial \phi^2}-24a\left(B\omega a^{-3\omega}+i\frac{\partial}{\partial T}\right)\right]\psi(a,\phi,T)=0~.
\end{eqnarray}
This can be written in the form
\begin{eqnarray}
\hat{\mathcal{H}}\psi=i\frac{\partial}{\partial t}\psi
\end{eqnarray}
where
\begin{eqnarray}
\hat{\mathcal{H}}=\frac{\partial^2}{\partial a^2}-\frac{6}{F(\phi)a^2}\frac{\partial^2}{\partial \phi^2}-24a\left(B\omega a^{-3\omega}\right)~.
\end{eqnarray} 
To solve this equation we now make the following ansatz\footnote{Note that $T=t$ corresponds to the time coordinate in the above equation.}
\begin{equation}
\psi(a,\phi,t)=e^{iEt}\Phi(a,\phi)~.
\end{equation}
This gives
\begin{eqnarray}
\label{4b}
\frac{\partial^2\Phi(a,\phi)}{\partial a^2}-\frac{6}{F(\phi)a^2}\frac{\partial^2\Phi(a,\phi)}{\partial \phi^2}-24a\left(B \omega a^{-3\omega}-E\right)\Phi(a,\phi)=0~.
\end{eqnarray}
We note that for $\hat{\mathcal{H}}$ to be a self-adjoint operator, the inner product between any two wave functions $\psi_1$ and $\psi_2$ must satisfy
\begin{eqnarray}
\label{4bb}
(\psi_1, \psi_2)=\int_{(a,\phi)}aF(\phi)\psi_1^*\psi_2 da d\phi
\end{eqnarray}
with the boundary conditions
\begin{eqnarray}
\label{4g}
\psi(0,\phi,t)=\psi(a,0,t)=\frac{\partial\psi(a,\phi,t)}{\partial a}\bigg|_{a=0}=\frac{\partial\psi(a,\phi,t)}{\partial \phi}\bigg|_{\phi=0}=0~.
\end{eqnarray}
Note that the inner product given in \cite{B. Majumder} does not make $\hat{\mathcal{H}}$ self-adjoint since the operator $\partial_\phi$ acts on $F(\phi)$ appearing in the denominator of the $\phi$-term in the Hamiltonian.
 
\noindent Applying the method of separation of variables once again by setting
\begin{equation}
\Phi(a,\phi)=\eta(a)\zeta(\phi)
\end{equation}
leads to the following
\begin{eqnarray}
\label{4c}
\frac{a^2(t)}{\eta(a)}\frac{d^2\eta(a)}{da^2}-24a^3(B\omega a^{-3\omega}-E) &=& -\kappa^2\\
\label{4d}
\frac{d^2\zeta(\phi)}{d\phi^2}+\frac{\kappa^2}{6}F(\phi)\zeta(\phi) &=& 0
\end{eqnarray}
where $\kappa^2$ is the separation constant.
Solving eq.(\ref{4c}) for $\omega=1$, we get
 
\begin{eqnarray}
\label{4e}
\eta(a)&=&\sqrt{a} \sqrt{2}\left(\frac{E}{3}\right)^\frac{1}{6} \left[C'_1\Gamma(1-r) J_{-r}\left(4\sqrt{\frac{2E}{3}}a^{\frac{3}{2}}\right)+C'_2\Gamma(1+r) J_r\left(4\sqrt{\frac{2E}{3}}a^{\frac{3}{2}}\right)\right] \\ \nonumber
&=&\sqrt{a}\left[C_1 J_{-r}\left(4\sqrt{\frac{2E}{3}}a^{\frac{3}{2}}\right)+C_2 J_r\left(4\sqrt{\frac{2E}{3}}a^{\frac{3}{2}}\right)\right] 
\end{eqnarray}
where 
\begin{eqnarray}
r&=&\frac{1}{3}\sqrt{1+96B-4\kappa^2}\\ \nonumber
C_1&=&C'_1\Gamma(1-r)\sqrt{2}\left(\frac{E}{3}\right)^\frac{1}{6}\\ \nonumber
C_2&=&C'_2\Gamma(1+r)\sqrt{2}\left(\frac{E}{3}\right)^\frac{1}{6}~.
\end{eqnarray}
Note that the above solution reduces to an Airy function if $96B-4\kappa^2 =0$.
This is the solution that one gets in the dust dominated universe ($\omega=0$)
in \cite{Pedram et. al} with Chaplygin gas and perfect fluid coupled to gravity. 

\noindent Considering $F(\phi)=6\lambda\phi^m,( m\neq-2$, $\lambda \textgreater 0)$, 
we get 
\begin{eqnarray}
\label{4f}
\nonumber \zeta(\phi)&=&(\kappa\lambda)^\frac{1}{m+2}(m+2)^\frac{-1}{m+2}\sqrt{\phi}\left[C'_3\Gamma(l-1) J_{-l}\left(\frac{2\kappa\lambda}{m+2}\phi^{\frac{m+2}{2}}\right)+C'_4\Gamma(l+1) J_{l}\left(\frac{2\kappa\lambda}{m+2}\phi^{\frac{m+2}{2}}\right)\right] \\ 
&=&(\kappa\lambda)^\frac{1}{m+2}(m+2)^\frac{-1}{m+2}\sqrt{\phi}\left[C_3J_{-l}\left(\frac{2\kappa\lambda}{m+2}\phi^{\frac{m+2}{2}}\right)+C_4J_{l}\left(\frac{2\kappa\lambda}{m+2}\phi^{\frac{m+2}{2}}\right)\right]
\end{eqnarray}
where
\begin{eqnarray}
l=\frac{1}{m+2},~
C_3=C'_3\Gamma(l-1),~
C_4=C'_4\Gamma(l+1).
\end{eqnarray}
We observe that the boundary conditions are satisfied if we set $C_1 = 0 = C_3$. The wave function therefore becomes
\begin{eqnarray}
\label{4h}
\psi(a,\phi,t)=C_2C_4(\kappa\lambda)^\frac{1}{m+2}(m+2)^\frac{-1}{m+2}\sqrt{a\phi}e^{iEt} J_{l}\left(\frac{2\kappa\lambda}{m+2}\phi^{\frac{m+2}{2}}\right)J_{r}\left(4\sqrt{\frac{2E}{3}}a^{\frac{3}{2}}\right)~.
\end{eqnarray}
With the above solution in hand, we now construct a wave packet by superposing all the eigenfunctions. We do this by first integrating over all possible values of $E$
\begin{eqnarray}
\label{4jj}
\psi_\kappa=C_5 \kappa^{\frac{1}{m+2}}\sqrt{a\phi}J_{l}\left(\frac{2\kappa\lambda}{m+2}\phi^{\frac{m+2}{2}}\right)\Gamma(1+r)\int^{\infty}_{\epsilon=0} A(\epsilon) e^{\frac{3\epsilon^2}{32}t}\epsilon^{\frac{4}{3}}J_{r}(\epsilon a^{\frac{3}{2}})d\epsilon ,~~~ \epsilon = 4\sqrt{\frac{4E}{3}}
\end{eqnarray}
where $A(\epsilon)$ is a weight factor that needs to be chosen properly so that the integration can be performed. Taking $A(\epsilon)=e^{-\gamma\epsilon^2}\epsilon^{r-\frac{1}{3}}$, we obtain
\begin{eqnarray}
\label{4ff}
\psi_\kappa=C_5 \kappa^{\frac{1}{m+2}}\sqrt{a\phi}J_{l}\left(\frac{2\kappa\lambda}{m+2}\phi^{\frac{m+2}{2}}\right)\Gamma(1+r)a^{\frac{3r}{2}}e^{\frac{-8a^3}{32\gamma-3it}}\left(2\gamma-\frac{3it}{16}\right)^{-(1+r)}~.
\end{eqnarray}
The next step to construct the wave packet is to integrate $\psi_\kappa$ over all possible values of $\kappa$. This yields
\begin{eqnarray}
\psi_{wp}&=&C_5 \sqrt{a\phi} e^{\frac{-8a^3}{32\gamma-3it}} \int_{\kappa=0}^{\frac{1}{4}+24B}  G(\kappa) \kappa^{\frac{1}{m+2}}J_{l}\left(\frac{2\kappa\lambda}{m+2}\phi^{\frac{m+2}{2}}\right)a^{\frac{3r}{2}} \left(2\gamma-\frac{3it}{16}\right)^{-(1+r)} \Gamma(1+r) d\kappa \nonumber
\end{eqnarray} 
where $G(\kappa)$ is a weight factor. Choosing $G(\kappa) = \kappa$, yields
\begin{eqnarray}
\label{4gg}
\psi_{wp}&=&C_5 \frac{m+2}{2\lambda}\phi^{-\frac{1}{2}(m+1)}\left(\frac{1}{4}+24B\right)^{1+\frac{1}{m+2}}a^{\frac{1}{2}(1+3\sqrt{1+96B})}e^{\frac{-8a^3}{32\gamma-3it}}\left(2\gamma-\frac{3it}{16}\right)^{-(1+\sqrt{1+96B})} \nonumber \\ &\times & J_{1+\frac{1}{m+2}}\left(\frac{1}{4}(1+96B)2\lambda\phi^{\frac{m+2}{2}}\right)~.
\end{eqnarray}
We now plot the probability densities at two different times for two different values of $B$. From the plots, we observe that the height of the peaks increases when the value of $B$ increases. A small increase in the value of $B$ results in  a large increase in the peak heights. Further, it can be easily seen that the height of the peaks decreases with the increase in time $T$.

\includegraphics[width=9cm]{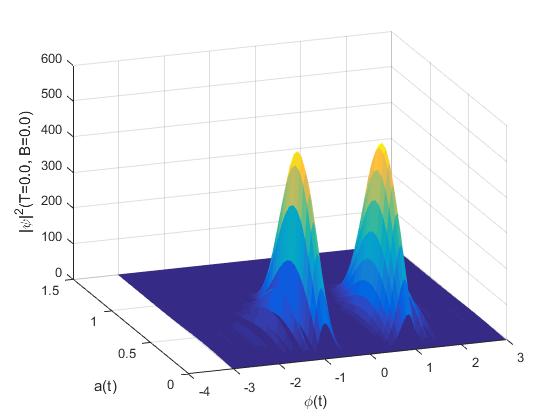}
\includegraphics[width=9cm]{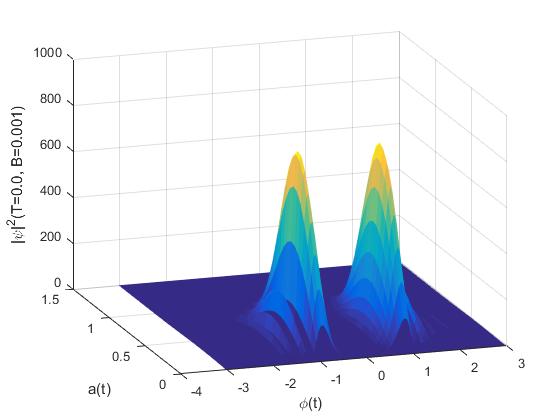}\\
\includegraphics[width=9cm]{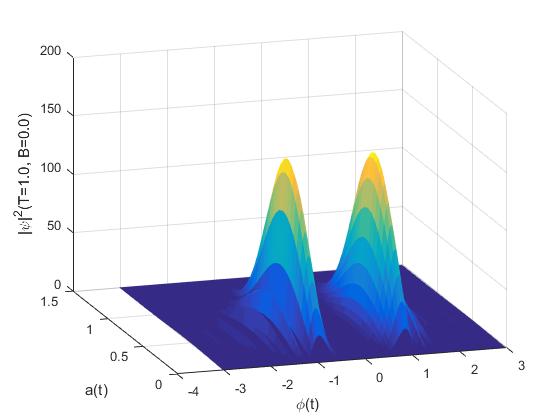}
\includegraphics[width=9cm]{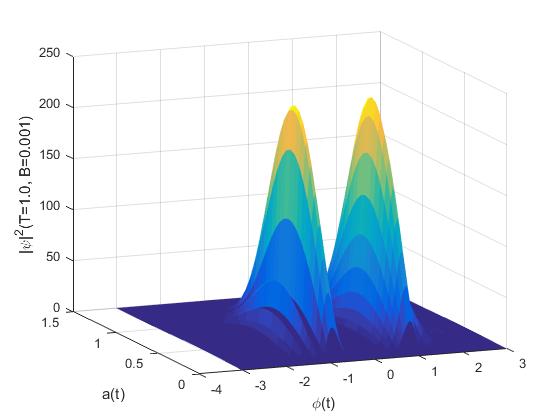}\\
\textsl{The plots show the behaviour of the probability density function, that is $|\psi_{wp}(a,\phi,T)|^2$ for two different time with two different values of $B$.}  \\~\\
Using the above quantum wave packet, we now calculate the expectation values of $a(t)$ and $\phi(t)$. Using the definition of the inner product between two wave functions (\ref{4bb}), we have
\begin{eqnarray}
\label{5a}
\langle a \rangle(t)=\frac{\int^\infty_{a=0}\int^\infty_{\phi=-\infty}aF(\phi)\psi^*_{wp} ~ a ~ \psi_{wp}dad\phi}{\int^\infty_{a=0}\int^\infty_{\phi=-\infty}aF(\phi)\psi^*_{wp}\psi_{wp}dad\phi}~.
\end{eqnarray}
Substituting eq.(\ref{4gg}) in the above relation yields 
\begin{eqnarray}
\label{5b}
\langle a \rangle(t)=\frac{1}{2}\left[16\gamma+\frac{9t^2}{64\gamma}\right]^{\frac{1}{3}}\frac{\Gamma\left(\frac{4}{3}+\sqrt{1+96B}\right)}{\Gamma\left(1+\sqrt{1+96B}\right)}~.
\end{eqnarray}
From the above result, we observe that the expectation value of the scale factor gets affected due to the presence of the perfect fluid.

\noindent The expectation value for the field $\phi$ reads
\begin{eqnarray}
\langle \phi \rangle(t)=\frac{\int^\infty_{a=0}\int^\infty_{\phi=-\infty}aF(\phi)\psi^*_{wp} ~ \phi ~ \psi_{wp}dad\phi}{\int^\infty_{a=0}\int^\infty_{\phi=-\infty}aF(\phi)\psi^*_{wp}\psi_{wp}dad\phi}
\end{eqnarray}
which gives 
\begin{eqnarray}
\langle \phi \rangle(t) = \textit{Constant}.
\end{eqnarray}

\section{Conclusion}
In this paper, we have studied the quantum cosmology of a scalar field coupled to a flat FRW spacetime in the presence of a generalized Chaplygin gas and perfect fluid. We observe that the inclusion of the perfect fluid in the matter sector has interesting consequences.
We have followed the Schutz's formalism to deal with the Chaplygin gas sector of the theory. The full theory is then quantized using the Wheeler-DeWitt approach. The Wheeler-DeWitt equation is solved for $\omega = 1$ using appropriate boundary conditions. The solution shows that there exists a choice of the constant appearing in the density of the perfect fluid for which the solutions reduce to the $\omega = 0$ solutions appearing in \cite{Pedram et. al}. The wave packet is then constructed from the solution of the Wheeler-DeWitt equation and exhibits two peaks as in \cite{B. Majumder}. The height of these peaks gets enhanced due to the presence of the perfect fluid.

\section*{Acknowledgment}
SG acknowledges the support by DST SERB under Start Up Research Grant (Young Scientist), File No.YSS/2014/000180. SG also acknowledges the support of IUCAA, Pune for the Visiting Associateship.

\end{document}